# Low-intensity illumination for lensless digital holographic microscopy with minimized sample interaction

**Bartosz Mirecki,[1,3] Mikołaj Rogalski,[1,3] Piotr Arcab,[1,3] Piotr Rogujski,[2] Luiza Stanaszek,[2] Michał Józwik[1] and Maciej Trusiak[1,*]**

[1]*Warsaw University of Technology, Institute of Micromechanics and Photonics, 8 Sw. A. Boboli St., 02-525 Warsaw, Poland*
[2]*NeuroRepair Department, Mossakowski Medical Research Institute, Polish Academy of Sciences, 5 Adolfa Pawinskiego St., 02-106 Warsaw, Poland*
[3]*Authors contributed equally to this work.*
**maciej.trusiak@pw.edu.pl*

**Abstract:** Exposure to laser light alters cell culture examination via optical microscopic imaging techniques, also based on label-free coherent digital holography. To mitigate this detrimental feature, researchers tend to use a broader spectrum and lower intensity of illumination, which can decrease the quality of holographic imaging due to lower resolution and higher noise. We study the lensless digital holographic microscopy (LDHM) ability to operate in the low photon budget (LPB) regime to enable imaging of unimpaired live cells with minimized sample interaction. Low-cost off-the-shelf components are used, promoting the usability of such a straightforward approach. We show that recording data in the LPB regime (down to 7 μW of illumination power) does not limit the contrast nor resolution of the hologram phase and amplitude reconstruction compared to the regular illumination. The LPB generates hardware camera shot noise, however, to be effectively minimized via numerical denoising. The ability to obtain high-quality, high-resolution optical complex field reconstruction was confirmed using the USAF 1951 amplitude sample, phase resolution test target, and finally, live glial restricted progenitor cells (as a challenging strongly absorbing and scattering biomedical sample). The proposed approach based on severely limiting the photon budget in lensless holographic microscopy method can open new avenues in high-throughout (optimal resolution, large field-of-view and high signal-to-noise-ratio single-hologram reconstruction) cell culture imaging with minimized sample interaction.

## 1. Introduction

Digital holographic microscopy (DHM) [1] is a widely used technique that allows for recording and reconstructing (via digital holographic principle [2]) an optical field that has been modulated via scattering, refraction, absorption, and reflection, by a biomedical [3] or technical [1] micro-object. Complex data at the registration plane can be obtained via hologram demodulation using one of the following single-frame carrier-fringes techniques with inclined reference beam: Fourier Transform [4], spatial carrier phase shifting [5] or Hilbert Transform [6-7]; it also can be performed by a more accurate multi-frame technique, e.g., temporal phase shifting [1,2]. Reconstructed complex data carry information about amplitude (absorptive features) and phase (refractive features), which are the fundamental components for label-free quantitative imaging. In the last two decades quantitative phase imaging emerged as one of the top label-free live bio-specimen examination frameworks [8,9]. After reconstruction it is possible, using, e.g., the angular spectrum (AS) method [10-14], to numerically propagate the optical field to the focus plane, if the hologram was captured outside of it. The focus plane is often determined automatically using so-called autofocusing approaches [15-20], contemporarily also performed within deep learning frameworks [21].

Quantitative phase microscopes implementing DHM principles, as well as classical bright-field microscopes, are bulky and consist of several optical elements (such as microscope objectives, mirrors, lenses, and gratings). In lensless digital holographic microscopy (LDHM) [22-25], a technology based on the seminal Gabor in-line holography concept [26], the system comprises of a light source, a CCD camera, and a sample placed between them, preferably near the CCD matrix [23-25]. This configuration determines the large field of view (FOV) primarily dependent on the size of the CCD matrix. In this case the resolution is mainly limited by the size of the pixel, which needs to be small enough to capture dense Gabor holographic fringes without any optical magnification [27]. The total number of pixels sampling the hologram is also crucial – the higher the number the larger the available information bandwidth which also affects resolution (both in terms of reconstructed details and captured object-scattered angles). The contrast of fringes, which in LDHM is mainly a function of spatial and temporal coherence of light source [27], is also crucial for the reconstruction resolution. The too low contrast of dense fringes prevents transferring detailed object features. Depth-of-focus, FOV and resolution are not limited by the microscope objective, as it is not required in LDHM (additionally free from objective induced aberrations). For these reasons LDHM gained significant attention in live-cell *in vitro* and tissue whole-slide bio-imaging [28-40], and technical object examination, i.e., micro/nano object tracking [41,42] (to name only some applications).

When using laser light, the problem of phototoxicity naturally arises. It is far more dangerous in fluorescent imaging [43], especially in techniques based on nonlinear effects [44] or super-resolution [43]. Lens-based DHM techniques, using both commercially available devices and custom setups, enabled minimally invasive time-lapse live cell imaging for quantification of cell growth, motility, and morphology alterations, see for example [45-50] and references therein. Potential photodamage can be considered as a likelihood in label-free microscopy, however, as mainly highly coherent laser light is used and once tightly structured or focused (e.g., optical tweezers [51], multiphoton label-free microscopy [52]), shorten in pulse duration [53] or illuminated onto a sample in a combination of long exposure time (time-lapse studies) and high intensity (required for high contrast of the hologram due to optical losses) laser VIS radiation can be appraised as sample-altering [54]. Especially in the case of high-energy short-wavelength illumination [55-58] for long-exposure time-lapse studies, as the UV light tends to damage the DNA of live cells [59]. Interestingly, other label-free microscopy techniques, e.g., Brillouin microscopy [60] or coherent anti-Stokes Raman microscopy [61] are also burdened by possible photodamage. To limit the potential sample interaction and damage low light intensity is used, referred to as low photon budget LPB imaging. It is also recommended to use broader spectra, however decreased temporal coherence can reduce the resolution of holographic imaging. Low temporal coherence also imposes stringent requirements in terms of available optical path difference (OPD) range and affects hologram recording in classical DHM (often requiring additional grating for OPD compensation [62-67]) and somewhat less critically, however still significantly, in LDHM setups (although no additional components are needed the imaging resolution is limited [69-70]). Therefore, it is potentially advantageous and desirable to use laser illumination with a short wavelength to realize high-throughput LDHM imaging (high lateral resolution and single-shot reconstruction for large FOV) without living cell impairment. A low photon budget is thus required to attain low-intensity illumination for lensless digital holographic microscopy with minimized live biological sample interaction.

Imaging in LPB has been studied within the interference fringe pattern based DHM framework [71-73] both in terms of assessing the noise level and proposing ways to remove it, e.g., upon deep learning based hologram-hologram transformation [73] and angular matching method [72]. Novel techniques have also been proposed to enhance the signal-to-noise ratio in phase imaging performed in LPB via the transport of intensity method [74,75] and deep-learning-aided phase retrieval schemes [76-78]. Low photons count related shot noise influence has been successfully studied in the DHM regime [79,80], however such an investigation is lacking for the LDHM which inspired this work. We plan to fill this gap and carefully examine the impact the LPB regime has on the LDHM phase/amplitude imaging capabilities. In LDHM the lack of optical elements positions it as an excellent tool for imaging biological specimens using very low intensity of light source illumination. Recording data in LPB regime (lowering the output power of the laser) with a given camera exposure time generates hardware shot noise, which can be identified by a Poisson noise [71-77]. The presence of these defects also affects the signal-to-noise ratio (SNR) of the hologram reconstruction: both phase and amplitude terms of the reconstructed complex optical field in the object plane are somewhat degenerated. This hardware noise can be decreased in single-camera-shot only by the numerical image processing methods, e.g., Gaussian filter, median filter or BM3D filter [81], due to the LPB regime.

In this contribution we propose to conduct LDHM imaging in the LPB regime, to obtain high-quality, high-resolution optical field reconstruction minimizing light-sample interaction and decreasing the camera shot noise. The aim is to corroborate LDHM as a promising tool for LPB operation and to promote it for live cell time-lapse inspection. To process reconstructed holograms, i.e., their shot-noise-affected phase and amplitude components, we used a well-established numerical Block-matching and 3D filtering (BM3D) algorithm, because of its ability to reduce noise and simultaneously keep the edges, fine features and dimensions of the objects. The BM3D is one of the expansions of the non-local means methodology [82]. There are two steps in BM3D: a hard-thresholding and a Wiener filter stage, both involving the following parts: grouping, collaborative filtering, and aggregation. The "strength" of the BM3D filtering depends mainly on the sigma parameter, which denotes the standard deviation of the Gaussian noise estimated to be present in the image. The ability to obtain high-quality, high-resolution optical field reconstruction was confirmed using (1) USAF 1951 as amplitude sample, (2) especially manufactured phase resolution test, and (3) living biological cells. We selected glial restricted progenitors as challenging biomedical sample to investigate in the LDHM system in the LPB regime. They are to be studied in very long time-lapses (up to several weeks) to investigate their migration schemes upon various stimuli. The challenging nature of these cells comes from their strong absorption and scattering (mainly amplitude objects) and the need to use specialized dish with dedicated coatings for the cells to grow, which additionally increases the scattering and thus augments the noise in hologram reconstructions. Studying migration of rather large cells (20-30 μm) we are not strongly restricted by rather low resolution (around pixel size [27] – in our case 2.4 μm) offered by the LDHM (especially when comparing to the commercial DHM devices). Instead, we take advantage of the extremely large field of view provided by the LDHM. All results corroborate easy-to-obtain high-quality LDHM imaging in photon starved conditions. This unique feature of LDHM is highlighted for the first time to the best of our knowledge.

The manuscript is constructed as follows. Section 2 describes the optical setup used and presents amplitude test target and phase test target LDHM imaging results with varying laser output power and various numerical denoising techniques used. Section 3 contains the quantitative analysis of the BM3D-driven amplitude and phase maps denoising in LPB LDHM. Section 4 shows experimental validation of the proposed low photon budget LDHM imager employed for live cells time-lapse investigation (in comparison with regular laser output power). Section 5 concludes the paper. Presented analysis, both qualitative and quantitative (estimating resolution using phase/amplitude targets and noise levels via calculating standard deviation) constitute a valuable experimental study of LDHM capabilities in LPB. Moreover, it provides important evidence corroborating the ability to perform accurate amplitude/phase LDHM imaging under LPB, which is very advantageous for live cell in-vitro biomedical imaging with minimized sample-interaction. We have also shown that Poisson shot noise is the main limiting factor and it can be efficiently minimized upon numerical filtration of reconstructed phase and amplitude maps.

## 2. Limiting the laser output power in the LDHM

The LDHM setup used within this study is presented in Fig. 1. Deployed lensless setup is a classical system to record in-line Gabor holograms, with a modification to reduce the intensity of the violet laser beam (CNI Lasers MDL-III-405-20mW, λ = 405 nm, FWHM = 23 pm). Rotating linear polarizer located between the laser light source and microscopic objective is used to reduce the intensity and control the low photon budget regime. Light is then coupled via microscope objective (x20/0.4) to the single-mode fiber. Due to its narrow core (fi 2.5 – 3.4 μm) it provides high-quality spherical beam (which works similarly to the pinhole). The sample is localized near to the CMOS matrix (ALVIUM Camera 1800 U-2050m mono Bareboard, pixel size 2.4 x 2.4 μm, 5496x3672 pixels, FOV = 116 $mm^2$) to obtain a high NA and scattered light collection capacity. Due to the camera's protected glass this distance is equal to approximately 2 mm. The magnification of this setup is around M = 1, because of the significant distance (300 mm) between the point source and CMOS.

The lensless digital holographic microscope allows to capture large FOV which is shown in Fig. 1(a). We depict an entire FOV in-line hologram of the USAF 1951 test target. Numerical propagation of recorded hologram to the object focus plane enables obtaining high-resolution image of the amplitude, Fig. 1(b), and phase, Fig. 1(c), samples. Yellow rectangles mark the areas of interest used to calculated noise (as the standard deviation of the object-free area [83]).

To calculate the in-focus optical field the object wave $u_{z=0}$, registered at plane $z$ = 0 is propagated to a defined distance $z$, found via an automatic focusing algorithm [16], which can be done using, e.g., angular spectrum method [10-13]:

$$\tilde{u}_{z=0}\left(f_x, f_y\right) = \iint u_{z=0}\left(x, y\right)\exp\left[-i2\pi\left(f_x x + f_y y\right)\right]dxdy, \tag{1}$$

$$\tilde{u}_z\left(f_x, f_y\right) = u_{z=0}\left(f_x, f_y\right)\exp\left[-i2\pi z\sqrt{\left(\frac{n}{\lambda}\right)^2 - \left(f_x + f_y\right)^2}\right], \tag{2}$$

$$u_z\left(x, y\right) = \iint \tilde{u}_z\left(f_x, f_y\right)\exp\left[i2\pi\left(f_x x + f_y y\right)\right]df_x df_y. \tag{3}$$

In Eqs. (1)-(3) tilde denotes Fourier transform, $i$ is an imaginary unit, ($f_x$,$f_y$) are spatial frequencies, $\lambda$ stands for the light wavelength, n is the refractive index of the medium in which the propagation takes place.

The purpose of the proposed experiment is an examination of captured lensless LPB hologram. To change the intensity of the beam a polarizer was rotated to different angle positions. Reducing the illumination power coming from the optical fiber and maintaining a constant exposure time of the recorded images cause a hardware-based shot-noise to appear. Figure 2(a) presents phase and amplitude noise (calculated as the standard deviation of the background object-free regions marked in Figs. 1(b) and 1(c)) plots estimated within reconstructed holograms recorded employing output laser power range: 7 μW - 50 μW. Micro Watt regime constitute a low photon budget; up to now sub milli Watts were used in the literature [73]. Noise makes it challenging to recognize and distinguish the details of the reconstructed holograms, both in amplitude Fig. 2(b) and phase Fig. 2(c). To counteract it, we can significantly extend the exposure time, which generally has two undesirable effects: (1) it limits the temporal resolution of the imaging device and (2) it increases the light-sample interaction needed to yield a single image. Other solution is to use appropriate filtering of the obtained hologram. Due to the quantitative metrological nature of the LDHM, the optimal choice would be to use a filter that does not distort the reconstructed amplitude and phase components and maintains the features of the examined objects.

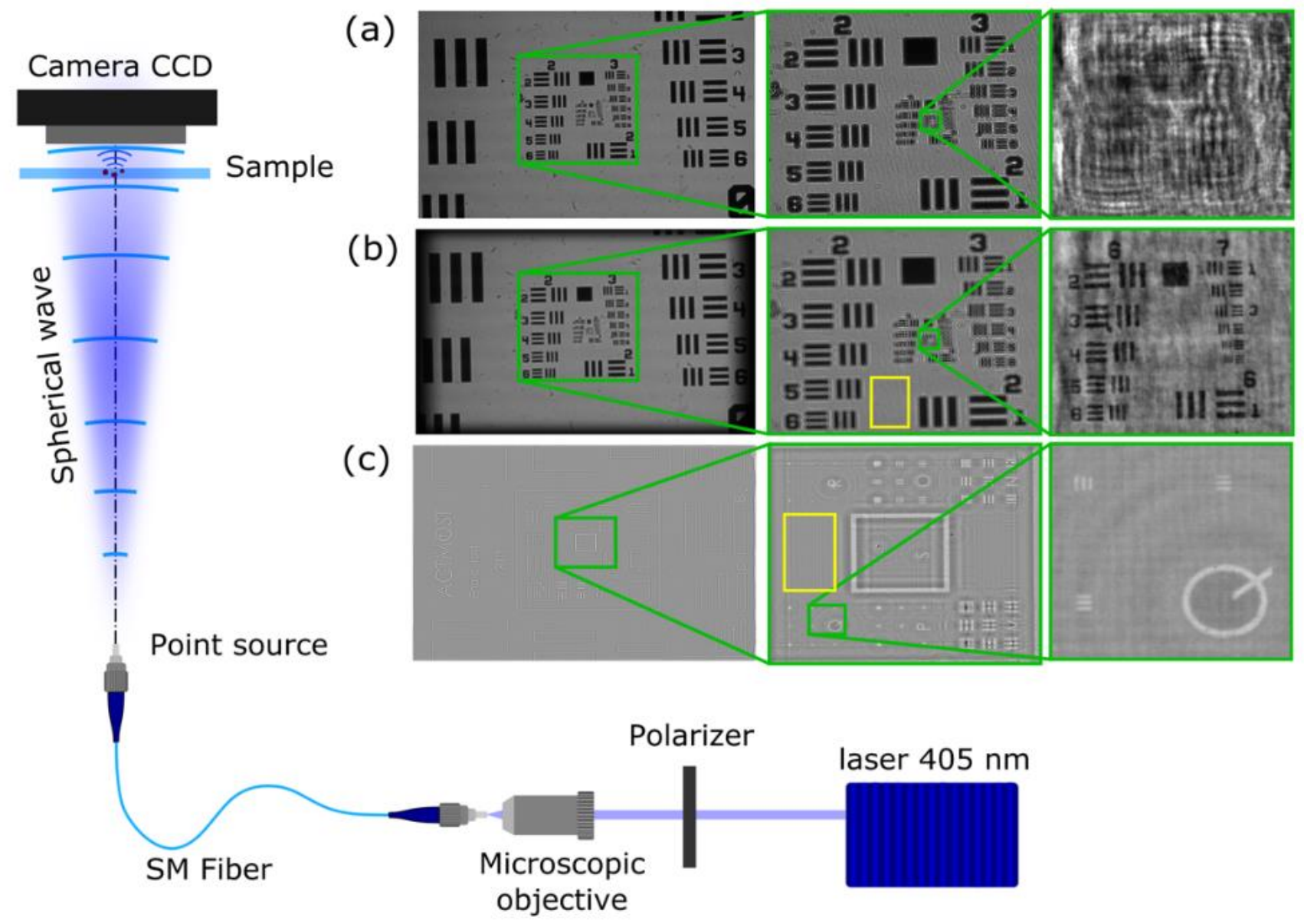

Fig. 1. Schematic layout of the LDHM setup. Exemplifying recorded and reconstructed images: (a) hologram of USAF amplitude test target, (b) amplitude term of the reconstructed hologram (a), (c) phase term of the reconstructed hologram of the phase test target. Regions of interest containing fine features have been enlarged for all images. Yellow rectangles encapsulate object-free areas utilized throughout the paper to estimate the amplitude and phase noise levels by calculating the standard deviation parameter.

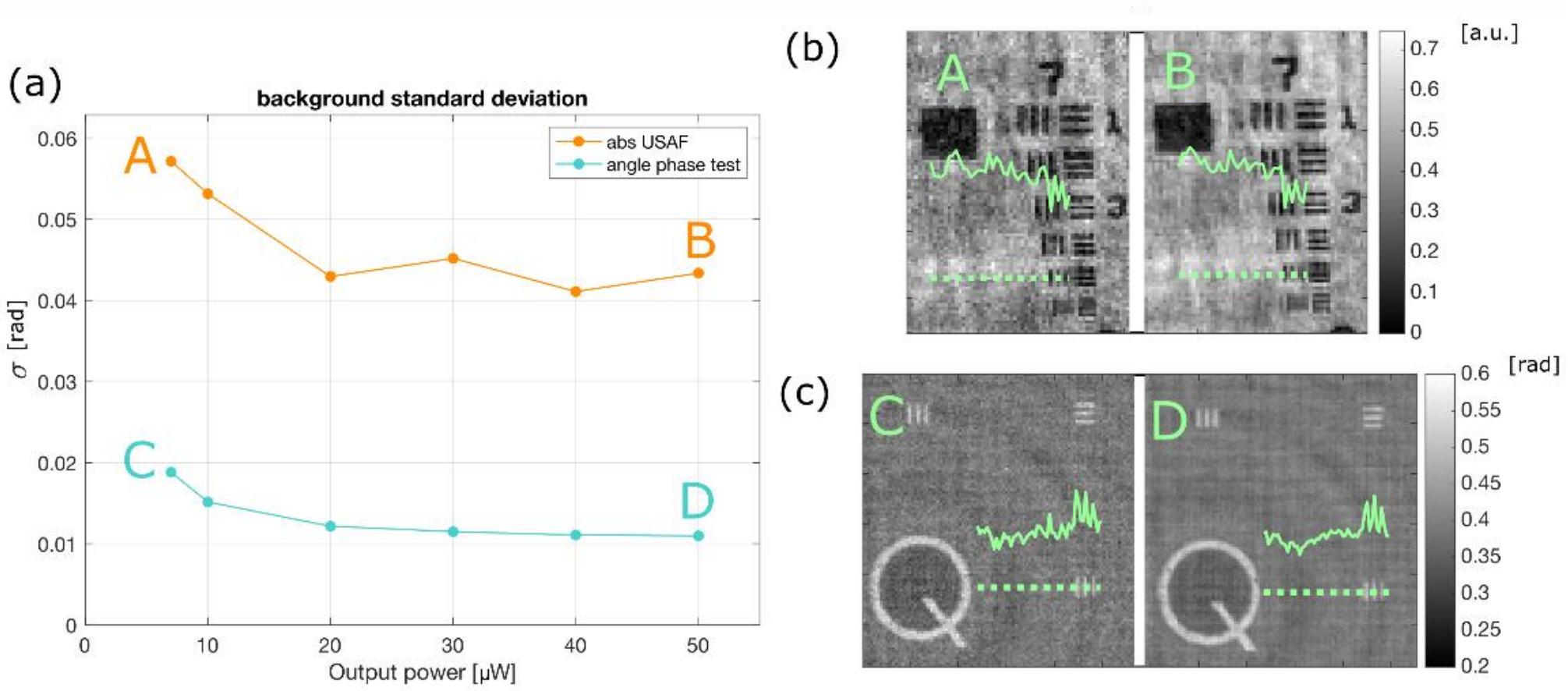

Fig. 2. (a) Plots of background standard deviation versus laser output power, for (orange) amplitude term of the USAF 1951 hologram reconstruction and (blue) phase term of the phase test target hologram reconstruction, (b) enlargements of the USAF 7th group with cross-sections for 7 μW (A) and 50 μW (B) illumination power, (c) enlargements of the phase target with cross-sections for 7 μW (C) and 50 μW (D) illumination power.

The robustness of the LDHM technique to the LPB regime stems from the fact that in the in-line Gabor configuration the information from a single pixel in the object plane is encoded within a large area in the hologram plane containing of defocused Gabor fringes (overlapping area of defocused object-scattered field and not-scattered light). This robust spatial way of encoding the object information is additionally merged with angular spectrum method for numerical backpropagation, which uses the entire spectrum of the Gabor hologram. Fourier transforms of Gabor holograms recorded under regular illumination and low photon budget regime are very similar, with majority of the information encoded in the central part with overlapped autocorrelation and cross-correlation terms. In a off-axis DHM [73] there is a need to transform a fringe pattern into the complex valued field, where only a small part of the detector bandwidth is used (centered around the carrier spatial frequency lobe) and the LPB introduces strong fringe intensity noise which is transferred to the phase map [73]. Although Fourier transform phase demodulation method is a global one, the physics behind off-axis in-plane hologram formation used in the DHM is a local pixel-to-pixel one, without the beneficial features of areal spatial information encoding in the defocused Gabor regime characteristic to the LDHM.

## 3. Noise minimization without jeopardizing the amplitude and phase resolution in LPB LDHM

As camera shot noise is deteriorating the quality of the LDHM reconstructions in the LPB we investigate the methods for its minimization. Figure 3(a) shows standard deviation plots calculated from the background (object-free) part of the reconstructed holograms marked in the Fig. 1. We employed three different amplitude and phase map filtering methods: BM3D, Gaussian filter, and median filter. The filter coefficients values have been selected to obtain similar background standard deviation values in hologram reconstructions: median – 3x3, gaussian – 5x5 with sigma 0.9, BM3D – sigma 0.003. This procedure allows for a global (single STD valued) quantitative comparison of the filtering methods. It suggests that Gaussian filtering provides the lowest noise level, however one needs to further evaluate this by assessing the details and important object features perseverance after denoising.

Figure 3(b) shows that Gaussian and median filters are ineffective because of the introduced significant blur. For median filtering resolution drops from 2.46 μm to 3.91 μm half-pitch. The BM3D is characterized by a good transition of the object features and allowed obtaining background noise minimization with resolution maintained as 2.46 μm half-pitch. Figure 4 presents cross-sections through the smallest resolvable element from USAF amplitude test target to allow for evaluation of the resolution in various filtering schemes. One can clearly observe that the BM3D filter preserved the resolution of the original un-filtered amplitude reconstruction. The BM3D is a collaborative filtering process which expands the non-local means methodology [82]. There are two steps in BM3D: a hard-thresholding and a Wiener filter stage, both involving the following parts: grouping, collaborative filtering, and aggregation. The “significance” of the BM3D filtering depends mainly on the sigma parameter, which denotes the standard deviation of Gaussian noise estimated to be present in the image (to be denoised).

To visualize and analyze the use of the BM3D filter in low photon budget LDHM, reconstructions were calculated for different illumination conditions and varying filter sigma coefficient. Sigma describes the estimated standard deviation of Gaussian noise present in the image data [81]. Our tests were performed for the USAF 1951 amplitude target, Fig. 5, and the phase test target, Fig. 6. We based our amplitude resolution analysis on the element 5 from group 7 of the USAF test with 2.46 μm half-pitch matching the camera pixel size of 2.4 μm. Phase test target has slightly larger Q elements of 3.5 μm half-pitch. It is important to notice that the resolution in x and y directions differs due to the rectangular camera matrix. Along the longer side we could capture more Gabor fringes, thus increasing the resolution.

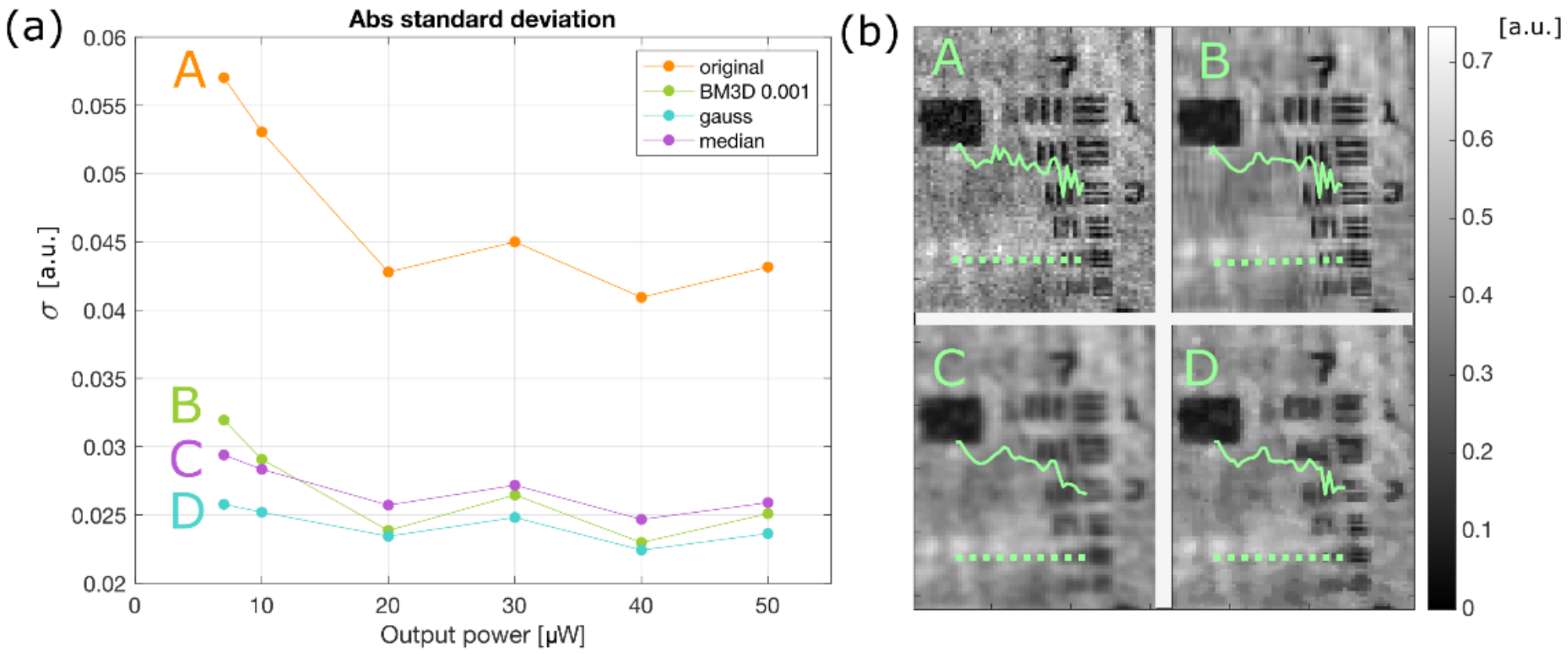

Fig. 3. (a) Plots of the background standard deviation versus the laser output power, for original and filtered USAF 1951 holograms amplitude reconstructions, (b) A-D enlargements of the 7th group with cross-sections for original, BM3D, Gauss and median filtered cases, respectively.

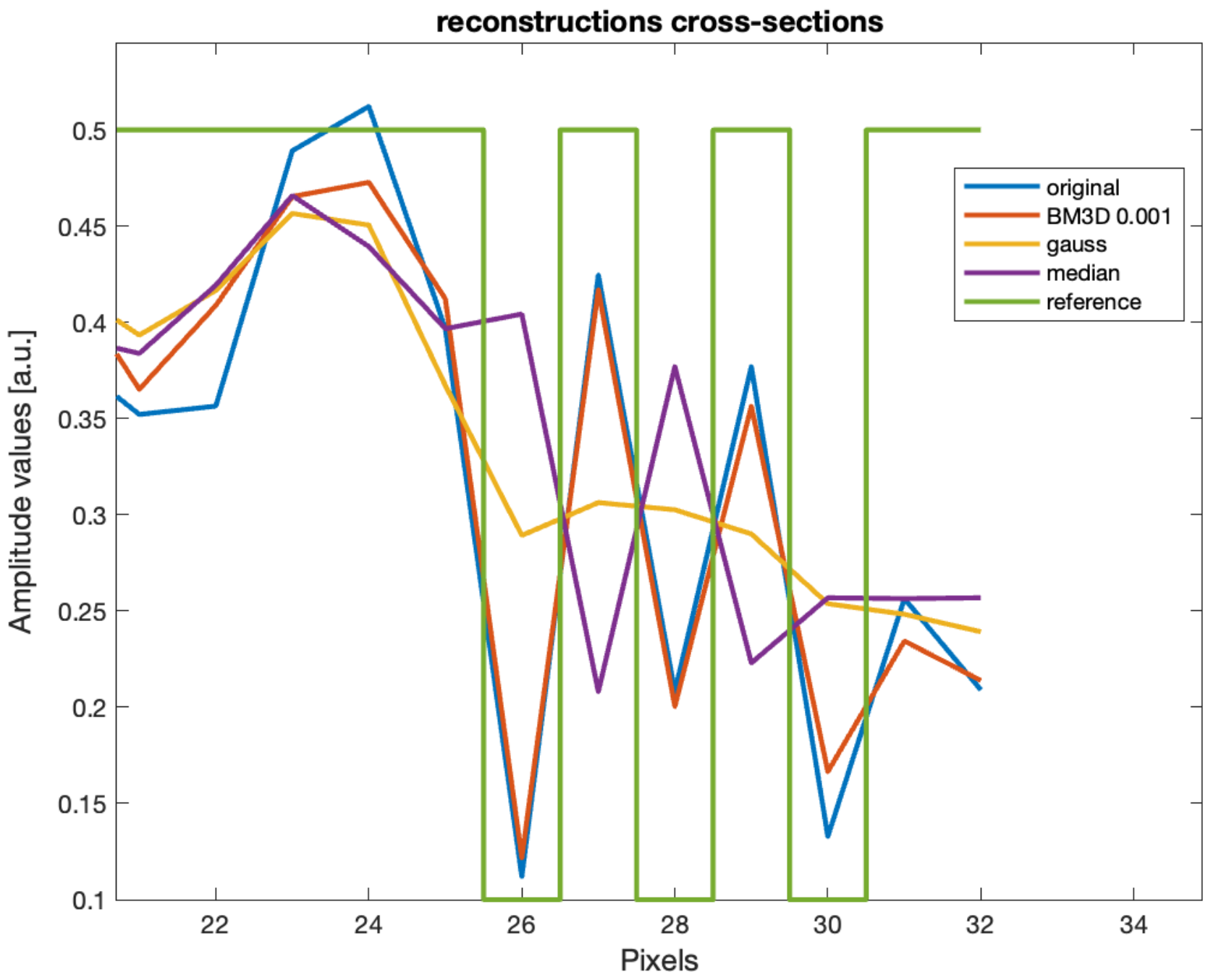

Fig. 4. Cross-sections through the 5th element from the 7th group of the USAF 1951 amplitude test target reconstructed with various filtering schemes.

Plots presented in Figs 5 and 6 show that after the BM3D filtering there is no need to increase the illumination power. Moreover, further increase of BM3D sigma value does not negatively affect the resolution. It generally means that we can obtain a better outcome with a filtered reconstructed image from a low photon budget hologram than unfiltered holograms acquired with higher power up to 50 µW (please compare, e.g., Fig. 5(b): D and C). Additionally, the resolution of the imaged objects is preserved, which can be observed analyzing cross-sections showed in Figs. 7 and 8. The presented discussion has similar outcomes for amplitude and phase LDHM imaging; however, phase noise is generally of lower standard deviations. It is also crucial to note, that holograms recorded using laser output power equal to 7 μW have a very limited dynamic range, as only 5 different gray levels are seen in histograms.

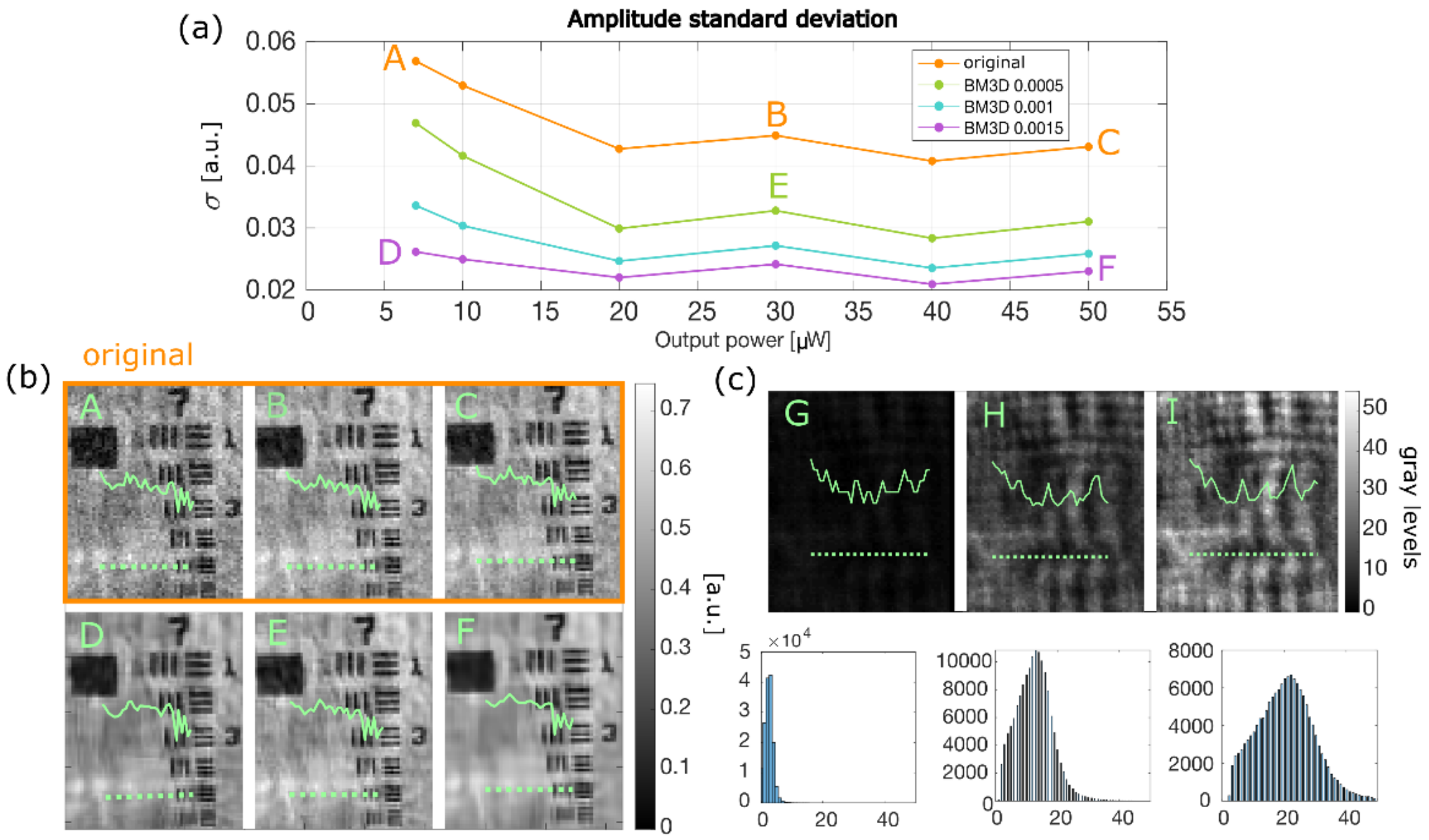

Fig. 5. (a) Plots of the background standard deviation versus the laser output power for original and filtered USAF 1951 hologram amplitude reconstructions, (b) A-F enlargements of the 7th group with cross-sections for original and filtered reconstructions marked within the plots presented in Fig. 5(a), (c) G-I enlargements of non-reconstructed holograms with cross-sections and histograms which correspond to the A-C reconstructions, respectively. Histogram x-axis denotes the gray level value and y-axis denotes the number of pixels.

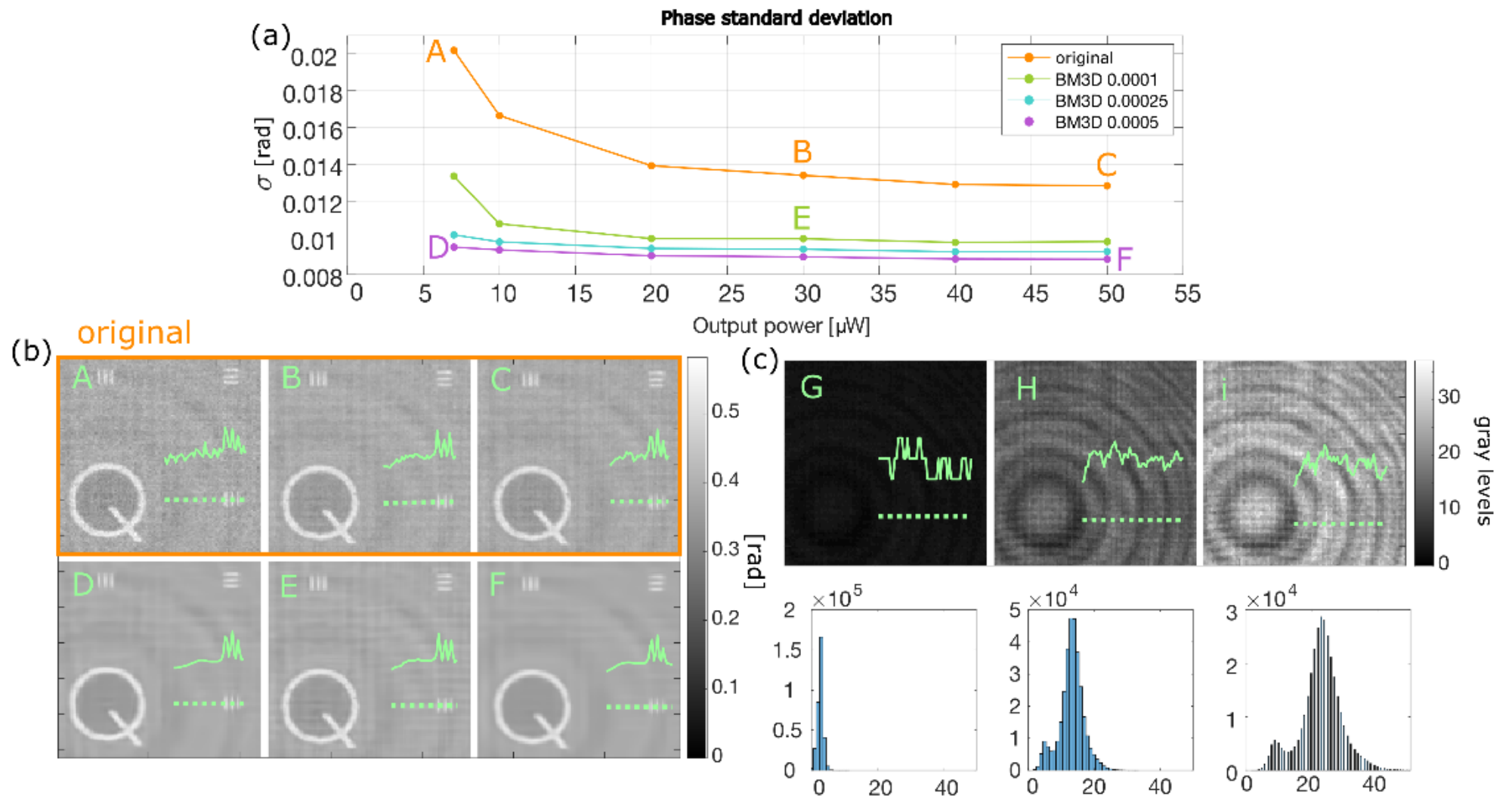

Fig. 6. (a) Plots of the background standard deviation versus the laser output power for original and filtered customized phase test target hologram phase reconstructions, (b) A-F enlargements of the Q element with cross-sections for original and filtered reconstructions marked within the plots presented in Fig. 6(a), (c) G-I enlargements of non-reconstructed holograms with cross-sections and histograms which correspond to the A-C reconstructions, respectively. Histogram x-axis denotes the gray level value and y-axis denotes the number of pixels.

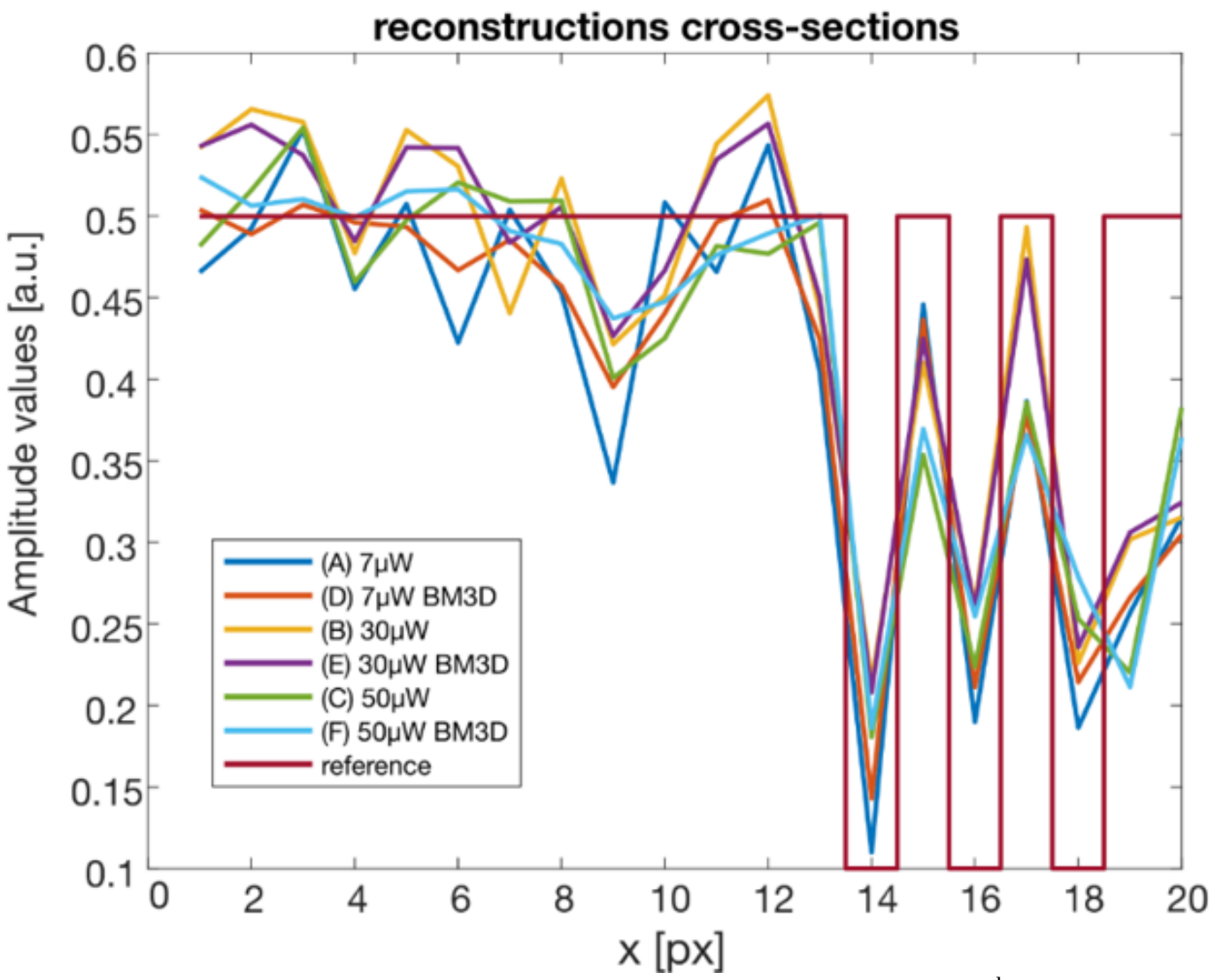

Fig. 7. Cross-sections for reference, original and filtered amplitude features (the 5th element from the 7th group of the USAF 1951 amplitude test target) obtained upon reconstruction of holograms recorded with varying illumination power.

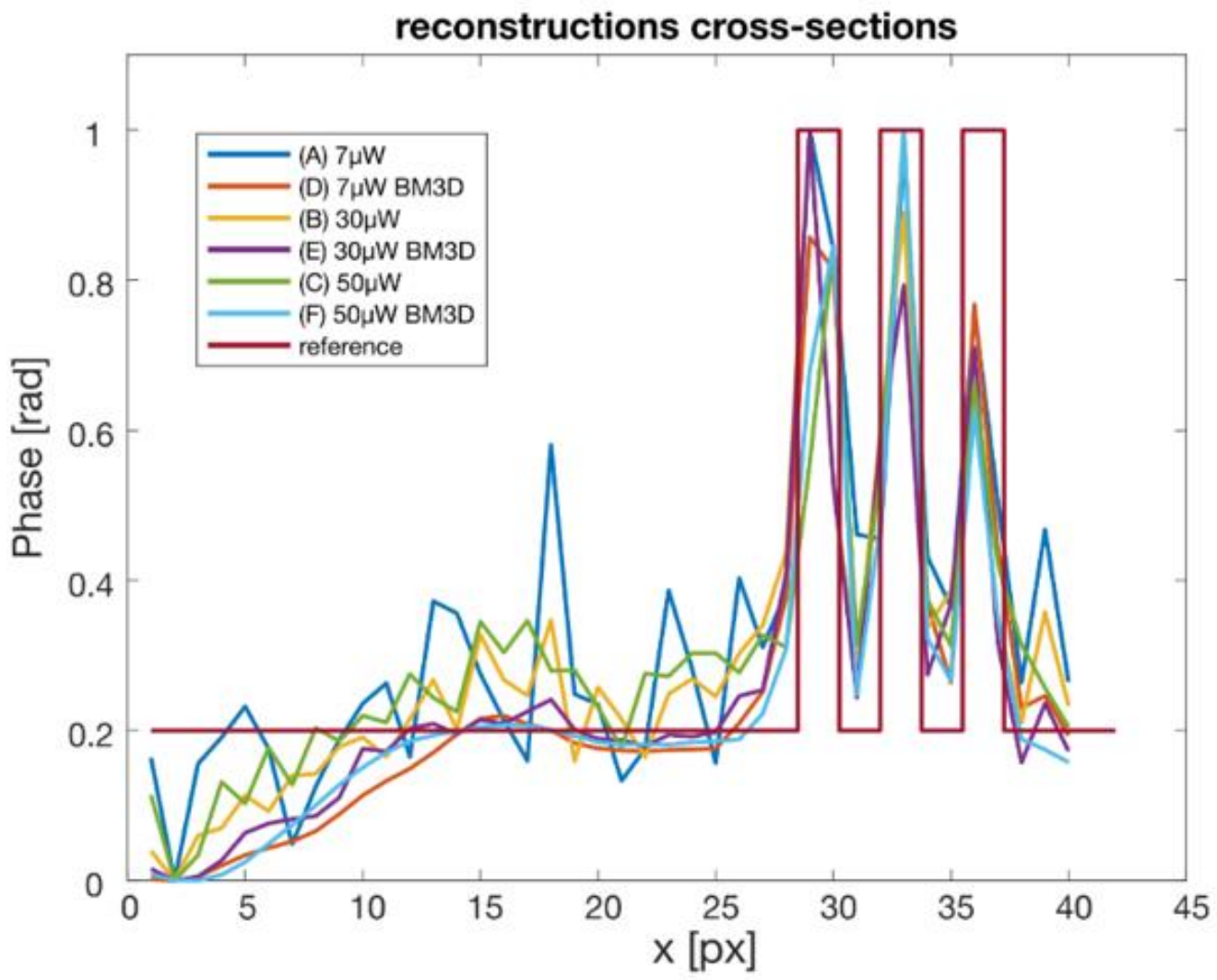

Fig. 8. Cross-sections for reference, original and filtered phase features (element from the Q-group of the customized phase test target) obtained upon reconstruction of holograms recorded with varying illumination power.

## 4. Experimental results for live cell imaging

We selected glial restricted progenitors (GRPs) as challenging biomedical sample to investigate in LDHM system. GRPs are the primary cells that exhibit the potential to differentiate into astrocytes and oligodendrocytes. Both types of cells closely cooperate with neurons to ensure their proper functioning. Due to their crucial role in the physiology and pathology of the central nervous system, a stem cell research area is recently focused on the biology of GRPs especially their key features, like proliferation, migration and differentiation. Therefore, the development of new methods of cell visualization allowing for the observation of their in vitro behavior is of key importance. GRPs were isolated from mouse brain fetuses (E13, embryonic day 13th) as described elsewhere [84]. Cells were cultured in a standard culture conditions at 37 °C, 21% $O_2$, 5% $CO_2$ until 80% confluence and passaged into the 35 mm glass-bottom dish with a 20 mm micro-well (Cellvis) and visualized using our LDHM system.

The cells we have used are interesting in terms of application in stem cell therapy. Therefore, the data concerning their migratory capacity are of great value. Even though interesting from the application point of view, cells are not easy to culture. The GRPs we work on have high requirements when it comes to culturing surface. Not only they need to be cultured on the special PE-treated flask/dishes, but also the dish surface needs to be coated with poly-L-lysine and laminin. Thus, in order to perform experiments with the use of the LDHM we have to take into consideration high scattering of the sample substrate. Moreover, the cells strongly absorb and scatter the light themselves, which increases the troublesome nature of performed experiments. Additionally, the cells are to be studied in a very long time-lapses (up to several weeks) to investigate their migration schemes upon various stimuli, thus limiting the energy via LPB LDHM is crucial to not alter the experiments [54]. Studying migration of rather large cells (20-30 μm) we are not strongly restricted by rather low resolution (around pixel size – in our case it is 2.4 μm) offered by the LDHM (especially when comparing to the commercial DHM devices). Instead, we took advantage of the extremely large field of view provided by the LDHM and significantly increase the space-time-bandwidth product of cell migration sensing.

Due to very challenging nature of the sample resulting in noisy reconstructions we modified the numerical reconstruction path. Previously we used simple angular spectrum method as static phase and amplitude test samples were of high quality. In the experimental live cell imaging, we subtract from each hologram the mean frame (from the sequence of grabbed holograms) to remove noise coming from static artifacts and imperfections. After that we reconstruct the background-removed holograms with iterative Gerchberg-Saxton [85] method employing non-negativity constraints [86,87].

Reconstruction has been performed initially for the regular illumination, i.e., laser output power high enough to record holograms with a full dynamic range of the camera. Then we minimized the laser output power to 7 μW and recorded further migration of the living cell colony. Figures 9(a) and 9(b) present exemplifying holograms from time-lapse studies performed within standard illumination and LPB regime. Red squares visible in Figs. 9(a) and 9(b) mark the regions of interest evaluated in detail in Figs. 9(c)-9(e), in terms of amplitude reconstruction capabilities for standard illumination and LPB with additional BM3D-based numerical denoising. Phase maps are not presented here as they do not contain valuable information because the sample is strongly absorbing and scattering and primarily acts like an amplitude object. Enlarged areas depicted in Figs. 9(f)-9(g) show the BM3D filtering effect – slight reduction of background noise without unwanted blur. Visualization 1 contains dynamic sequences of holograms and amplitude reconstructions computed for the time-lapse study of live GRPs bio-imaging via the LDHM technique in high (HPB) and low photon budget scenarios with additional BM3D filtering. The presented results corroborate the LDHM method for the LPB regime large FOV examination of the cell migration (please see the Visualization 1 depicting readily observable cell movement and interactions). Conducted BM3D-based denoising showed promise in minimization of the background noise.

To experimentally validate the proposed technique in quantitative phase imaging we performed measurements of fixed biological sample of neuronal cells from rodent brain. Mixed neuronal hippocampal cell cultures were obtained from P0 Wistar rats. Cultures were plated on coverslips in a 12-well plate in Neurobasal supplemented with B-27 (Invitrogen) and kept in controlled conditions (37C, 4% $CO_2$) as described previously [88]. Next, cells were fixed using 4% paraformaldehyde/4% sucrose for 10 mins at RT and washed with PBS. For imaging coverslips were placed on glass slides and mounted with Fluoromount. We captured lensless holograms using regular illumination and in low photon budget scenario. Figure 10 presents obtained phase reconstruction results: Fig. 10(a) shows full field of view reconstructed under regular illumination, Fig. 10(b) depicts enlarged area encapsulated within the red rectangle marked in Fig. 10(a), Fig. 10(c) illustrates the same area reconstructed in LPB regime (additional noise is observable), Fig. 10(d) displays the BM3D filtered LPB reconstruction (noise is successfully reduced). To quantify the difference in illumination scenarios, we presented histograms of regular and LPB holograms in Fig. 10(e) and Fig. 10(f), respectively. It is worth showcasing that despite a much smaller number of gray levels, see Fig. 10(f), the LPB regime reconstruction exhibits a similar quality of imaged cells as regular illumination scenario, differing only in greater Poisson noise. After employing the BM3D filtering, the results showed in Fig. 10(b) and Fig. 10(d) are visually very similar.

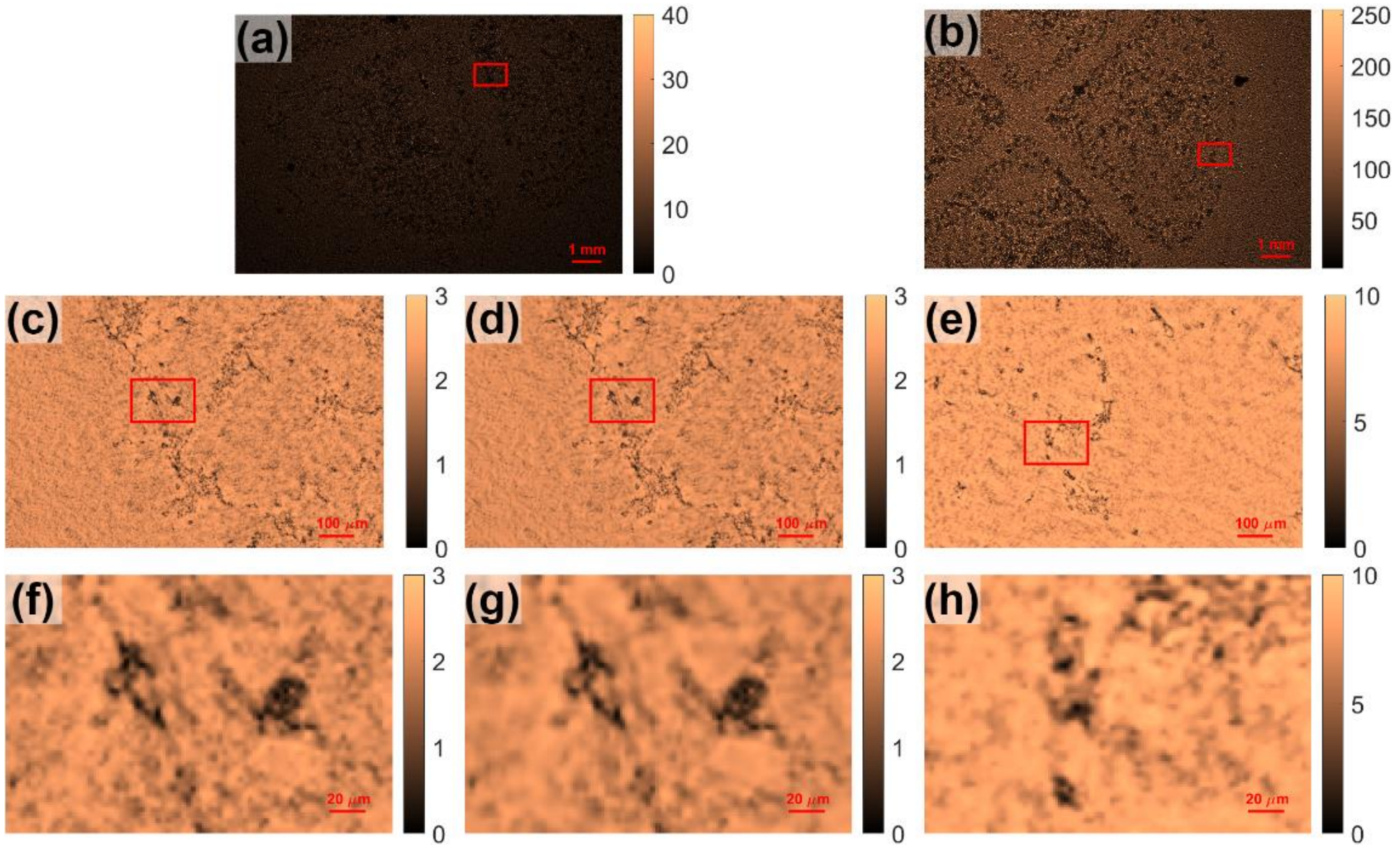

Fig. 9. Experimental evaluation via live GRPs imaging: exemplifying holograms recorded using (a) LPB regime and (b) regular illumination; low light conditions resulted in a very obscure LPB hologram Fig. 9(a), even in visualization range truncated from 0-255 to 0-40 gray levels. (c) enlarged area marked within red rectangle in Fig. 9(a) without and (d) with additional BM3D filtering; (e) enlarged red rectangle area marked in Fig. 9(b). (f)-(h) additionally zoomed-in regions marked by red rectangles in corresponding Figs. 9(c)-9(e). Visualization 1 contains dynamic sequences of amplitude reconstructions for low (LPB) and high (HPB) photon budget regimes (corresponding to Figs. 9(c)-9(e)).

## 5. Conclusions

We presented a study on the lensless digital holographic microscopy. We investigated its ability to operate in the low photon budget regime to enable imaging of unimpaired live cells with minimized sample interaction. Low-cost off-the-shelf components were used, which promotes the usability of such a straightforward approach. We corroborated that recording data in the LPB regime (down to 7 μW of illumination power) does not limit the contrast nor resolution of the hologram phase/amplitude reconstruction in comparison with the regular illumination (defined as providing full dynamic range usage of the recorded holograms in an 8-bit CMOS camera). As LPB generates hardware camera shot noise, we evaluated the BM3D as a competent algorithm to effectively minimize it. The ability to obtain large FOV, high-quality, high-resolution optical complex field reconstruction was confirmed using static calibrated samples: USAF 1951 amplitude test target and phase resolution test target. Live dynamic glial restricted progenitor cells constituted a challenging biomedical sample. The proposed approach based on severely limiting the photon budget in the lensless digital holographic microscopy method can open new avenues in high-throughout (optimal resolution, large field-of-view and high signal-to-noise-ratio single-hologram reconstruction) cell culture imaging with minimized sample interaction.

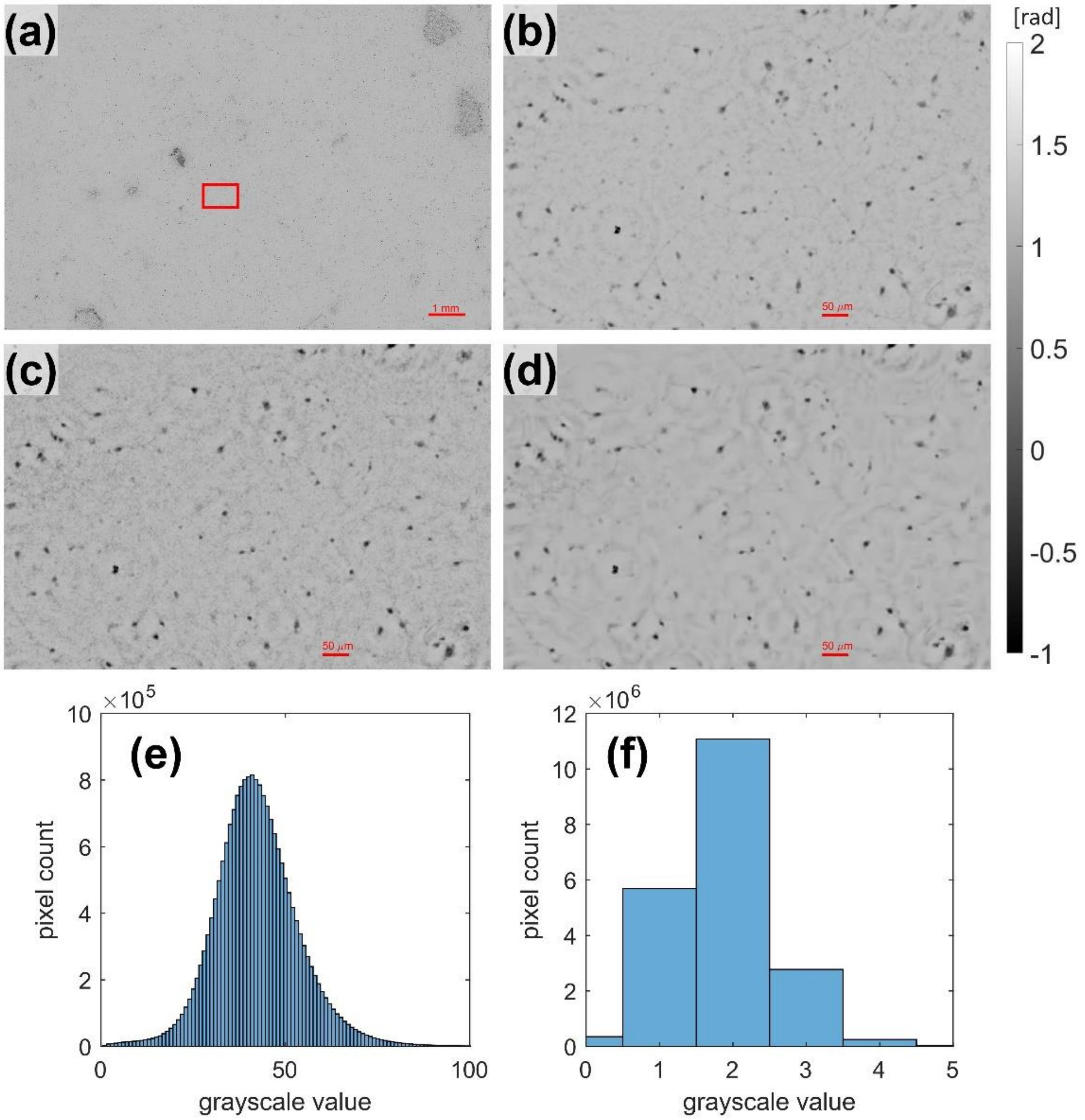

Fig. 10. Experimental imaging of neural cells. Phase reconstruction results: (a) full field of view reconstructed under regular illumination, (b) enlarged area encapsulated within the red rectangle marked in Fig. 10(a), (c) the same area reconstructed in LPB regime (additional noise is readily observable), (d) the BM3D filtered LPB reconstruction (noise presence is successfully reduced). To quantify the difference in illumination scenarios histograms of (e) regular and (f) LPB holograms are presented.

**Acknowledgements.** Authors would like to thank Marzena Stefaniuk, Monika Pawłowska and Aleksandra Mielnicka from the Nencki Institute of Experimental Biology (Polish Academy of Sciences) for providing a fixed neuronal cells sample analyzed in Fig. 10. MT thanks Professor Krzysztof Patorski for fruitful discussions.

**Funding.** This work has been funded by the National Science Center Poland (SONATA 2020/39/D/ST7/03236) and European Social Fund (POWR.03.02.00-00-I028/17-00).

**Disclosures.** The authors declare no conflicts of interest.

**Data availability.** Data underlying the results presented in this paper are not publicly available at this time but may be obtained from the authors upon reasonable request.

## References

1. M.K. Kim, *Digital Holographic Microscopy* (Springer, 2011).
2. U. Schnars and W. Jueptner, *Digital Holography* (Springer, 2005).
3. B. Kemper and G. von Bally, "Digital holographic microscopy for live cell applications and technical inspection," Appl. Opt. **47**(4), A52-A61 (2008).
4. M. Takeda, H. Ina, and S. Kobayashi, "Fourier-transform method of fringe-pattern analysis for computer-based topography and interferometry," J. Opt. Soc. Am. **72**(1), 156-160 (1982).
5. M. Pirga and M. Kujawinska, "Two directional spatial-carrier phase-shifting method for analysis of crossed and closed fringe patterns," Opt. Eng. **34**(8), (1995).
6. T. Ikeda, G. Popescu, R. Dasari, and M. Feld, "Hilbert phase microscopy for investigating fast dynamics in transparent systems," Opt. Lett. **30**(10), 1165-1167 (2005).
7. M. Trusiak, M. Cywińska, V. Micó, J. A. Picazo-Bueno, C. Zuo, P. Zdańkowski, and K. Patorski, "Variational Hilbert Quantitative Phase Imaging," Sci. Rep. **10**, 13955 (2020).
8. Y. Park, C. Depeursinge, and G. Popescu, "Quantitative phase imaging in biomedicine," Nat. Photonics **12**, 578–589 (2018).
9. G. Popescu, *Quantitative Phase Imaging of Cells and Tissues* (McGraw-Hill, 2011).
10. F. Shen and A. Wang, "Fast-Fourier-transform based numerical integration method for the Rayleigh-Sommerfeld diffraction formula," Appl. Opt. **45**(6), 1102-1110 (2006).
11. K. Matsushima and T. Shimobaba, "Band-Limited Angular Spectrum Method for Numerical Simulation of Free-Space Propagation in Far and Near Fields," Opt. Express **17**(22), 19662-19673 (2009).
12. T. Kozacki, K. Falaggis, and M. Kujawinska, "Computation of diffracted fields for the case of high numerical aperture using the angular spectrum method," Appl. Opt. **51**(29), 7080-7088 (2012).
13. T. Kozacki and K. Falaggis, "Angular spectrum-based wave-propagation method with compact space bandwidth for large propagation distances," Opt. Lett. **40**(14), 3420-3423 (2015).
14. K. M. Molony, B. M. Hennelly, D. P. Kelly, and T. J. Naughton, "Reconstruction algorithms applied to in-line Gabor digital holographic microscopy," Opt. Commun. **283**(6), 903–909 (2010).
15. P. Langehanenberg, G. von Bally, and B. Kemper, "Autofocusing in digital holographic microscopy," 3D Res **2**, 4 (2011).
16. F. Dubois, A. El Mallahi, J. Dohet-Eraly, and C. Yourassowsky, "Refocus criterion for both phase and amplitude objects in digital holographic microscopy," Opt. Lett. **39**(15), 4286-4289 (2014).
17. P. Memmolo, M. Paturzo, B. Javidi, P. A. Netti, and P. Ferraro, "Refocusing criterion via sparsity measurements in digital holography," Opt. Lett. **39**(16), 4719-4722 (2014).
18. M. Trusiak, J.-A. Picazo-Bueno, P. Zdankowski, and V. Micó, "DarkFocus: numerical autofocusing in digital in-line holographic microscopy using variance of computational dark-field gradient," Opt. Laser Eng. **134**, 106195 (2020).
19. C. A. Trujillo and J. Garcia-Sucerquia, "Automatic method for focusing biological specimens in digital lensless holographic microscopy," Opt. Lett. **39**(9), 2569-2572 (2014).
20. Y. Zhang, H. Wang, Y. Wu, M. Tamamitsu, and A. Ozcan, "Edge sparsity criterion for robust holographic autofocusing," Opt. Lett. **42**(19), 3824-3827 (2017).
21. Y. Wu, Y. Rivenson, Y. Zhang, Z. Wei, H. Günaydin, X. Lin, and A. Ozcan, "Extended depth-of-field in holographic imaging using deep-learning-based autofocusing and phase recovery," Optica **5**(6), 704-710 (2018).
22. J. Garcia-Sucerquia, W. Xu, S. K. Jericho, P. Klages, M. H. Jericho, and H. J. Kreuzer, "Digital in-line holographic microscopy," Appl. Opt. **45**(5), 836-850 (2006).
23. A. Greenbaum, W. Luo, T-W. Su, Z. Göröcs, L. Xue, and S. O. Isikman, "Imaging without lenses: achievements and remaining challenges of wide-field on-chip microscopy," Nat. Methods **9**, 89-95 (2012).
24. A. Ozcan and E. McLeod, "Lensless Imaging and Sensing," Annu. Rev. Biomed. Eng. **18**, 77–102 (2016).
25. E. McLeod and A. Ozcan, "Unconventional methods of imaging: computational microscopy and compact implementations," Rep. Prog. Phys. **79**, 076001 (2016).
26. D. Gabor, "A New Microscopic Principle," Nature **161**, 777–8 (1948).
27. T. Agbana, H. Gong, A. Amoah, V. Bezzubik, M. Verhaegen, and G. Vdovin, "Aliasing, coherence, and resolution in a lensless holographic microscope," Opt. Lett. **42**(12), 2271-2274 (2017).
28. C. Allier, S. Morel, R. Vincent, L. Ghenim, F. Navarro, M. Menneteau, T. Bordy, L. Hervé, O. Cioni, and X. Gidrol, "Imaging of dense cell cultures by multiwavelength lens-free video microscopy," Cytom. A **91**(5), 433–442 (2017).
29. M. Sanz, J. Á. Picazo-Bueno, L. Granero, J. García, and V. Micó, "Compact, cost-effective and field-portable microscope prototype based on MISHELF microscopy," Sci. Rep. **7**, 43291 (2017).
30. L. Herve, O. Cioni, P. Blandin, F. Navarro, M. Menneteau, T. Bordy, S. Morales, and C. Allier, "Multispectral total-variation reconstruction applied to lens-free microscopy," Biomed. Opt. Express **9**(11), 5828-5836 (2018).
31. Y. Rivenson, Y. Wu, and A. Ozcan, "Deep learning in holography and coherent imaging," Light Sci. Appl. **8**, 85 (2019).

32. W. Xu, M. Jericho, I. Meinertzhagen, and H. Kreuzer, "Digital in-line holography for biological applications," Proc. Natl. Acad. Sci. **98**(20), 11301–5 (2001).
33. A. B. Bochdansky, M. Jericho, and G. J. Herndl, "Development and deployment of a point-source digital inline holographic microscope for the study of plankton and particles to a depth of 6000 m," Limnol. Oceanogr.-Meth. **11**(1), 28-40 (2013).
34. T-W. Su, L. Xue, and A. Ozcan, "High-throughput lensfree 3D tracking of human sperms reveals rare statistics of helical trajectories," Proc. Natl. Acad. Sci. U.S.A. **109**(40), 16018-16022 (2012).
35. T-W. Su, I. Choi, J. Feng, K. Huang, and A. Ozcan, "High-throughput analysis of horse sperms' 3D swimming patterns using computational on-chip imaging," Anim. Reprod. Sci. **169**, 45-55 (2016).
36. J. Ryle, S. McDonnell, B. Glennon, and J. Sheridan, "Calibration of a digital in-line holographic microscopy system: depth of focus and bioprocess analysis," Appl. Opt. **52**(7), C78-C87 (2013).
37. R. Corman, W. Boutu, A. Campalans, P. Radicella, J. Duarte, M. Kholodtsova, L. Bally-Cuif, N. Dray, F. Harms, G. Dovillaire, S. Bucourt, and H. Merdji, "Lensless microscopy platform for single cell and tissue visualization," Biomed. Opt. Express **11**(5), 2806-2817 (2020).
38. S. Amann, M. von Witzleben, and S. Breuer, "3D-printable portable open-source platform for low-cost lens-less holographic cellular imaging," Sci. Rep. **9**, 11260 (2019).
39. Y. Wu and A. Ozcan, "Lensless digital holographic microscopy and its applications in biomedicine and environmental monitoring," Methods **136**, 4–16 (2018).
40. H. Zhu, S. Isikman, O. Mudanyali, A. Greenbaum, and A. Ozcan, "Optical imaging techniques for point-of-care diagnostics," Lab Chip **13**(1), 51-56 (2019).
41. P. Memmolo, L. Miccio, M. Paturzo, G. Di Caprio, G. Coppola, P. A. Netti, and P. Ferraro, "Recent advances in holographic 3D particle tracking," Adv. Opt. Photon. **7**(4), 713-755 (2015).
42. O. Mudanyali, E. McLeod, and W. Luo, "Wide-field optical detection of nanoparticles using on-chip microscopy and self-assembled nanolenses," Nat. Photonics **7**, 247–254 (2013).
43. S. Wäldchen, J. Lehmann, T. Klein, S. van de Linde, and M. Sauer, "Light-induced cell damage in live-cell superresolution microscopy," Sci. Rep. **5**, 15348 (2015).
44. A. Hopt and E. Neher, "Highly Nonlinear Photodamage in Two-Photon Fluorescence Microscopy," Biophys. J. **80**(4), 2029-2036 (2001).
45. K. M. Eder, A. Marzi, A. Barroso, S. Ketelhut, B. Kemper, and J. Schnekenburger, "Label-Free Digital Holographic Microscopy for In Vitro Cytotoxic Effect Quantification of Organic Nanoparticles," Cells **11**(4), 644 (2022).
46. D. Bettenworth, P. Lenz, P. Krausewitz, M. Brückner, S. Ketelhut, D. Domagk, and B. Kemper, "Quantitative Stain-Free and Continuous Multimodal Monitoring of Wound Healing In Vitro with Digital Holographic Microscopy," PLoS ONE **9**(9), e107317 (2014).
47. M. Rezaei, J. Cao, K. Friedrich, B. Kemper, O. Brendel, M. Grosser, M. Adrian, G. Baretton, G. Breier, and H.-J. Schnittler, "The expression of VE-cadherin in breast cancer cells modulates cell dynamics as a function of tumor differentiation and promotes tumor–endothelial cell interactions," Histochem. Cell Biol. **149**, 15–30 (2018).
48. O. Tolde, A. Gandalovičová, A. Křížová, P. Veselý, R. Chmelík, D. Rosel, and J. Brábek, "Quantitative phase imaging unravels new insight into dynamics of mesenchymal and amoeboid cancer cell invasion," Sci. Rep. **8**, 12020 (2018).
49. M. Hellesvik, H. Øye, and H. Aksnes, "Exploiting the potential of commercial digital holographic microscopy by combining it with 3D matrix cell culture assays," Sci. Rep. **10**, 14680 (2020).
50. S. Aknoun, M. Yonnet, Z. Djabari, F. Graslin, M. Taylor, T. Pourcher, B. Wattellier, and P. Pognonec, "Quantitative phase microscopy for non-invasive live cell population monitoring," Sci. Rep. **11**, 4409 (2021).
51. Á. Peña, B. Kemper, M. Woerdemann, A. Vollmer, S. Ketelhut, G. von Bally, and C. Denz, "Optical tweezers induced photodamage in living cells quantified with digital holographic phase microscopy," Proc. SPIE 8427, 84270A (2012).
52. R. Galli, O. Uckermann, E. Andresen, K. Geiger, and E. Koch, "Intrinsic Indicator of Photodamage during Label-Free Multiphoton Microscopy of Cells and Tissues," PLoS ONE **9**(10), e110295 (2014).
53. B. Talone, M. Bazzarelli, A. Schirato, F. Dello Vicario, D. Viola, E. Jacchetti, M. Bregonzio, M. T. Raimondi, G. Cerullo, and D. Polli, "Phototoxicity induced in living HeLa cells by focused femtosecond laser pulses: a data-driven approach," Biomed. Opt. Express **12**(12), 7886-7905 (2021).
54. M. Baczewska, P. Stępień, M. Mazur, W. Krauze, N. Nowak, J. Szymański, and M. Kujawińska, "Method to analyze effects of low-level laser therapy on biological cells with a digital holographic microscope," Appl. Opt. **61**(5), B297-B306 (2022).
55. G. Pedrini, F. Zhang, and W. Osten, "Digital holographic microscopy in the deep (193 nm) ultraviolet," Appl. Opt. **46**(32), 7829-7835 (2007).
56. A. Faridian, D. Hopp, G. Pedrini, U. Eigenthaler, M. Hirscher, and W. Osten, "Nanoscale imaging using deep ultraviolet digital holographic microscopy," Opt. Express **18**(13), 14159-14164 (2010).
57. M. U. Daloglu, A. Ray, Z. Gorocs, M. Xiong, R. Malik, G. Bitan, E. McLeod, and A. Ozcan, "Computational On-Chip Imaging of Nanoparticles and Biomolecules using Ultraviolet Light," Sci. Rep. **7**, 44157 (2017).
58. M. U. Daloglu, A. Ray, M. J. Collazo, C. Brown, D. Tseng, B. Chocarro-Ruiz, L. M. Lechuga, D. Cascio, and A. Ozcan, "Low-cost and portable UV holographic microscope for high-contrast protein crystal imaging," APL Photonics **4**, 030804 (2019).

59. S. Nakajima, L. Lan, S. Kanno, K. Yamamoto, A. P.M. Eker, A Yasui, "UV Light-induced DNA Damage and Tolerance for the Survival of Nucleotide Excision Repair-deficient," Human Cells **279**(45), P46674-46677 (2004).
60. M. Nikolić and G. Scarcelli, "Long-term Brillouin imaging of live cells with reduced absorption-mediated damage at 660nm wavelength," Biomed. Opt. Express **10**(4), 1567-1580 (2019).
61. Y. Fu, H. Wang, R. Shi, and J-X. Cheng, "Characterization of photodamage in coherent anti-Stokes Raman scattering microscopy," Opt. Express **14**(9), 3942-3951 (2006).
62. K. Patorski, P. Zdańkowski, and M. Trusiak, "Grating deployed total-shear 3-beam interference microscopy with reduced temporal coherence," Opt. Express **28**(5), 6893-6908 (2020).
63. P. Zdańkowski, J. Winnik, K. Patorski, P. Gocłowski, M. Ziemczonok, M. Józwik, M. Kujawińska, and M. Trusiak, "Common-path intrinsically achromatic optical diffraction tomography," Biomed. Opt. Express **12**(7), 4219-4234 (2021).
64. C. Park, K. Lee, Y. Baek, and Y. Park, “Low-coherence optical diffraction tomography using a ferroelectric liquid crystal spatial light modulator,” Opt. Express **28(**26), 39649-39659 (2020).
65. R. Guo, S. K. Mirsky, I. Barnea, M. Dudaie, and N. T. Shaked, "Quantitative phase imaging by wide-field interferometry with variable shearing distance uncoupled from the off-axis angle," Opt. Express **28**(4), 5617-5628 (2020).
66. V. Mico, C. Ferreira, Z. Zalevsky, and J. García, "Spatially-multiplexed interferometric microscopy (SMIM): converting a standard microscope into a holographic one," Opt. Express **22**(12), 14929-14943 (2014).
67. P. Kolman and R. Chmelík, "Coherence-controlled holographic microscope," Opt. Express **18**(21), 21990-22004 (2010).
68. P. Petruck, R. Riesenberg, and R. Kowarschik, "Partially coherent light-emitting diode illumination for video-rate in-line holographic microscopy," Appl. Opt. **51**(13), 2333-2340 (2012).
69. J. Garcia-Sucerquia, "Noise reduction in digital lensless holographic microscopy by engineering the light from a light-emitting diode," Appl. Opt. **52**(1), A232-A239 (2013).
70. L. Repetto, E. Piano, and C. Pontiggia, "Lensless digital holographic microscope with light-emitting diode illumination," Opt. Lett. **29**(10), 1132-1134 (2004).
71. M. Marim, E. Angelini, J.-C. Olivo-Marin, and M. Atlan, "Off-axis compressed holographic microscopy in low-light conditions," Opt. Lett. **36**(1), 79-81 (2011).
72. K. Inoue, A. Anand, and M. Cho, "Angular spectrum matching for digital holographic microscopy under extremely low light conditions," Opt. Lett. **46**(6), 1470-1473 (2021).
73. Z. Zhang, Y. Zheng, T. Xu, A. Upadhya, Y. J. Lim, A. Mathews, L. Xie, and W. M. Lee, "Holo-UNet: hologram-to-hologram neural network restoration for high fidelity low light quantitative phase imaging of live cells," Biomed. Opt. Express **11**(10), 5478-5487 (2020).
74. A. Gupta and N. Nishchal, "Low-light phase imaging using in-line digital holography and the transport of intensity equation," J. Opt. **23**(2), 025701 (2021).
75. A. Gupta, N. Nishchal, and P. Banerjee, "Transport of intensity equation based photon-counting phase imaging," OSA Contin. **3**(2), 236-245 (2020).
76. M. Deng, A. Goy, S. Li, K. Arthur, and G. Barbastathis, "Probing shallower: perceptual loss trained Phase Extraction Neural Network (PLT-PhENN) for artifact-free reconstruction at low photon budget," Opt. Express **28**(2), 2511-2535 (2020).
77. A. Goy, K. Arthur, S. Li, and G. Barbastathis, "Low Photon Count Phase Retrieval Using Deep Learning," Phys. Rev. Lett. **121**, 243902 (2018).
78. I. Kang, F. Zhang, and G. Barbastathis, "Phase extraction neural network (PhENN) with coherent modulation imaging (CMI) for phase retrieval at low photon counts," Opt. Express **28**(15), 21578-21600 (2020).
79. F. Charriére, B. Rappaz, J. Kühn, T. Colomb, P. Marquet, and C. Depeursinge, "Influence of shot noise on phase measurement accuracy in digital holographic microscopy," Opt. Express **15**(14), 8818-8831 (2007).
80. C. Remmersmann, S. Stürwald, B. Kemper, P. Langehanenberg, and G. von Bally, "Phase noise optimization in temporal phase-shifting digital holography with partial coherence light sources and its application in quantitative cell imaging," Appl. Opt. **48**(8), 1463-1472 (2009).
81. K. Dabov, A. Foi, V. Katkovnik, and K. Egiazarian, "Image denoising by sparse 3D transform-domain collaborative filtering," IEEE Trans. Image Process. **16**(8), 2080–2095 (2007).
82. A. Buades, “A non-local algorithm for image denoising,” IEEE Computer Vision and Pattern Recognition **2**, 60–65 (2005).
83. J. Dohet-Eraly, C. Yourassowsky, A. El Mallahi, and F. Dubois, "Quantitative assessment of noise reduction with partial spatial coherence illumination in digital holographic microscopy," Opt. Lett. **41**(1), 111–114 (2016).
84. A. Lyczek, A. Arnold, J. Zhang, J. T. Campanelli, M. Janowski, J. W.M. Bulte, and P. Walczak, "Transplanted human glial-restricted progenitors can rescue the survival of dysmyelinated mice independent of the production of mature, compact myelin," Exp. Neurol. **291**, 74-86 (2017).
85. R. W. Gerchberg and W. O. Saxton, “A practical algorithm for the determination of phase from image and diffraction plane pictures,” Optik **35**, 237–246 (1972).
86. F. Momey, L. Denis, T. Olivier, and C. Fournier, "From Fienup’s phase retrieval techniques to regularized inversion for in-line holography: tutorial," J. Opt. Soc. Am. A **36**(12), D62-D80 (2019).

87. T. Latychevskaia, "Iterative phase retrieval for digital holography: tutorial," J. Opt. Soc. Am. A **36**(12), D31-D40 (2019).
88. M. Roszkowska, A. Krysiak, L. Majchrowicz, K. Nader, A. Beroun, P. Michaluk, M. Pekala, J. Jaworski, L. Kondrakiewicz, A. Puścian, E. Knapska, L. Kaczmarek, and K. Kalita, "SRF depletion in early life contributes to social interaction deficits in the adulthood," Cell. Mol. Life Sci. **79**, 278 (2022).